\documentclass[pra,showpacs,twocolumn]{revtex4} \usepackage{latexsym}
\usepackage{amsfonts} \usepackage{amsmath,revsymb} \usepackage{youngtab}\usepackage{graphicx}

\newcommand{\kt}[1]{\ensuremath{|#1\rangle}}

\newcommand{\bk}[2]{\ensuremath{\langle #1|#2\rangle}}
\newcommand{\HS}{\mathcal{H}}

\newcommand{\SHS}{\mathcal{S}}

\newcommand{\KHS}{\mathcal{K}}

\newcommand{\ckt}[1]{\ensuremath{|#1\}}}

\newcommand{\cbk}[2]{\ensuremath{\{ #1|#2\}}}

\newcommand{\pp}[1]{\ensuremath{\lfloor #1 \rfloor }}

\begin{document}

\title{Spectroscopy for a Few Atoms Harmonically-Trapped in One Dimension}

\author{N.L.~Harshman\footnote{Electronic address: harshman@american.edu}}

\affiliation{Department of Physics\\
4400 Massachusetts Ave.\ NW\\ American University\\ Washington, DC 20016-8058}

\begin{abstract}
Spectroscopic labels for a few particles with spin that are harmonically trapped in one-dimension with effectively zero-range interactions are provided by quantum numbers that characterize the symmetries of the Hamiltonian: permutations of identical particles, parity inversion, and the separability of the center-of-mass. The exact solutions for the non-interacting and infinitely repulsive cases are reduced with respect to these symmetries. This reduction explains how states of single-component and multi-component fermions and bosons transform under adiabatic evolution from non-interacting to strong hard-core repulsion. These spectroscopic methods also clarify previous analytic and numerical results for intermediate values of interaction strength. Several examples, including adiabatic mapping for two-component fermionic states in the cases $N=3-5$, are provided.
\end{abstract}

\maketitle

The quantum system of a few particles in one dimension, confined by an external field and interacting via a short-range potential, has been a touchstone model  for nuclear, atomic, statistical, and mathematical physics since the 1930's. This system possesses enough symmetry and regularity to be analytically tractable in many cases, but also manifests a rich enough phenomenological structure to serve as physically significant model for realistic systems. This class of models has been a playground for studying the interplay of integrability and solvability, the emergence of universality in few-body phenomena, and the transition from few-body to many-body physics.

This article focuses specifically on the quantum system of $N$ particles of mass $m$ confined in a one-dimensional harmonic trap with frequency $\omega$ and interacting via an effectively zero-range potential. The Hamiltonian for the model is
\begin{equation}\label{hammy}
\frac{\hat{H}}{\hbar\omega} = \frac{1}{2} \sum_{i=1}^N \left( -\frac{\partial^2}{\partial x_i^2} + x_i^2 \right) + g \sum_{i<j} \delta(x_i - x_j),
\end{equation}
with the particle coordinates $x_i$ expressed in units of $\sqrt{\hbar/m\omega}$. This Hamiltonian has received intense attention recently because it serves as a model for neutral cold atoms with interactions tunable through the manipulation of Feshbach and confinement-induced resonances trapped in extremely prolate, approximately harmonic optical wells~\cite{olshanii_atomic_1998, bloch_many-body_2008, chin_feshbach_2010, guan_fermi_2013}. The model has been used to investigate few-body and many-body properties of one-dimensional trapped ultracold gases of bosons~\cite{deuretzbacher_evolution_2007, deuretzbacher_exact_2008, girardeau_wave_2010, girardeau_tonks-girardeau_2011, ernst_simulating_2011, brouzos_construction_2012, wilson_geometric_2013}, fermions~\cite{guan_exact_2009, ma_mathematical_2009, yang_ground_2009, girardeau_two_2010, guan_super-tonks-girardeau_2010, brouzos_two-component_2013, bugnion_ferromagnetic_2013, gharashi_correlations_2013, sowinski_few_2013, lindgren_fermionization_2013, cui_ground-state_2014, volosniev_strongly-interacting_2013}, and mixtures of bosons and fermions~\cite{girardeau_soluble_2007, fang_exact_2011, wang_density-functional_2012}. The so-called unitary limit $g \rightarrow \infty$ of hard-core repulsive interactions is known as the Tonks-Girardeau gas and has been realized in the laboratory with bosons~\cite{paredes_tonksgirardeau_2004,kinoshita_observation_2004}, as has the metastable super-Tonks-Girardeau limit $g \rightarrow -\infty$~\cite{haller_realization_2009}. Recent experiments in which the state of a few trapped, interacting fermions can be controlled and measured~\cite{serwane_deterministic_2011, zurn_fermionization_2012, zurn_pairing_2013, wenz_few_2013} have generated excitement that tunable fermi gases will allow simulation of solid-state phenomena such as the transition to ferromagnetism~\cite{bugnion_ferromagnetic_2013, volosniev_strongly-interacting_2013, cui_ground-state_2014,deuretzbacher_quantum_2013}.

The solutions of Hamiltonian (\ref{hammy}) form a complete basis for the spatial states of $N$ distinguishable particles, and these solutions can be symmetrized to account for identical bosons or fermions, whether single species or multi-component. An exact, complete set of solutions of (\ref{hammy}) are known when $N=2$ for all $g$~\cite{busch_two_1998, jonsell_interaction_2002, farrell_universality_2010}, and these solutions have become the benchmark for further theoretical investigations. For $N \geq 3$, a complete set of exact solutions are known only for the interaction-free case $g=0$ and for the hard-core limit $g\rightarrow \infty$~\footnote{The $g\rightarrow \infty$ solutions are also energy eigenstates at $g\rightarrow -\infty$, but they no longer form a complete basis due to the existence of bound states.}. 
In the $g=0$ case, the system is equivalent to the well-known isotropic oscillator in $N$ dimensions. In the $g\rightarrow \infty$ limit, $N!$ independent solutions can be constructed from each $g=0$ fermionic solution~\cite{deuretzbacher_exact_2008, fang_exact_2011} by generalizing the famous Bose-Fermi mapping~\cite{girardeau_relationship_1960}. This mapping has also been generalized to construct exact solutions at $g\rightarrow \pm \infty$ for multi-component fermions~\cite{guan_exact_2009, girardeau_two_2010}, multi-component bosons~\cite{deuretzbacher_exact_2008, girardeau_tonks-girardeau_2011} and boson-fermion mixtures~\cite{fang_exact_2011}.

Both sets of exact spatial solutions can be described using a unified spectroscopic framework. The spectroscopic labels correspond to the irreducible representations (irreps) of the total symmetry group $\mathrm{G}_N$ of the Hamiltonian. The group $\mathrm{G}_N$ is isomorphic to the direct product $\mathrm{S}_N\times\mathrm{Z}_2\times \mathrm{U}(1)$, where $\mathrm{S}_N$ is permutation symmetry, $\mathrm{Z}_2$ corresponds to parity and $\mathrm{U}(1)$ to contact symmetry of the one-dimensional harmonic oscillator in the separated center-of-mass variable. The symmetry group $\mathrm{G}_N$ holds for any interaction strength $g\in\mathbb{R}$, so $\mathrm{G}_N$ irrep labels serve as quantum numbers that are respected by the interparticle interaction for all values of $g$, not just for the exact limits.

In this article, I argue that these spectroscopic labels can be used to help implement symmetrization, understand the transformation of states under rapid or adiabatic changes in the interaction parameter, and improve approximation schemes for states in the intermediate interaction regime. Many previous works have exploited the symmetry and representation theory of $\mathrm{S}_N$ to construct symmetrized states, e.g.\ \cite{guan_exact_2009, ma_mathematical_2009, yang_ground_2009, fang_exact_2011,cui_ground-state_2014}, but this treatment differs in three substantial ways. First, the symmetrization and reduction into $\mathrm{G}_N$ irreps in the $g=0$ and $g \rightarrow \infty$ limits is facilitated by using the Jacobi hypercylindrical coordinate system, defined below. This coordinate system naturally incorporates the contact and dilation  symmetries (c.f. \cite{werner_unitary_2006-1,werner_unitary_2006}) of $\hat{H}$. Second, a method is used to construct spatial states with the right properties for symmetrizing multi-component spin systems that does not require explicitly constructing the spin state. Third, and perhaps most importantly, the group  $\mathrm{G}_N$ exhausts the symmetry of $\hat{H}$ (except for  $g = 0$ or $g \rightarrow \pm \infty$), whereas $\mathrm{S}_N$ does not. Therefore, for intermediate values of $g$, each possible set of spectroscopic labels uniquely identifies a distinct degenerate energy subspace, and each subspace has the dimension of the corresponding $\mathrm{G}_N$ irrep.

When $g=0$ and $g\rightarrow \infty$ the symmetry group of (\ref{hammy}) is larger than $\mathrm{G}_N$ but contains it as a subgroup. That means the degenerate energy eigenspaces are reducible into irreps of $\mathrm{G}_N$, and they are usually not simply reducible. The reduction problem for these two exact limits is solved in this article. Comparing the pattern of irreps of $\mathrm{G}_N$ at $g=0$ and $g\rightarrow \infty$ provides the generalization of the boson-fermion mapping to multicomponent cases or mixed symmetry. Under adiabatic evolution of the interaction strength, states follow this mapping. The results of this method agree with recent calculations by other methods~\cite{gharashi_correlations_2013, volosniev_strongly-interacting_2013,deuretzbacher_quantum_2013, cui_ground-state_2014, sowinski_few_2013}. Additionally, once this map is known, one can construct approximate solutions for intermediate interaction strengths from the extreme exact cases using variational methods (c.f.\ \cite{wilson_geometric_2013}). In a subsequent work I will show how this same spectroscopic framework also allows the  dynamics arising from rapid variations of the interaction parameter, or quenches, to be described accurately and efficiently.

Finally, the spectroscopic notation developed here clarifies results achieved by a variety of analytic and numerical approximation methods for intermediate values of $g$~\cite{ernst_simulating_2011, brouzos_construction_2012, wang_density-functional_2012, armstrong_many-particle_2012, harshman_symmetries_2012, brouzos_two-component_2013,  bugnion_ferromagnetic_2013, gharashi_correlations_2013, sowinski_few_2013, lindgren_fermionization_2013, wilson_geometric_2013,damico_three_2013}. These methods typically produce spectral graphs that are dense with crossing lines, but these lines can be assigned unique labels with this system. When the graphs are partitioned into $\mathrm{G}_N$ irreps, there are no level crossings. Exact diagonalization methods that exploit this spectroscopic system will require fewer basis states to get the same accuracy because there are no interaction matrix elements between states in different $\mathrm{G}_N$ irreps.

\section{Reduction by $\mathrm{G}_N$}

The total Hilbert space for $N$ particles with $k$ accessible spin components and interaction strength $g$ is constructed as the tensor product $\HS=\KHS^g\otimes\SHS$ of the spatial Hilbert space $\KHS^g\sim\mathrm{L}^2(\mathbb{R}^N)$ and the spin Hilbert space $\SHS\sim\mathbb{C}^{k^N}$. The Hamiltonian $\hat{H}$ (\ref{hammy}) acts as the identity on $\SHS$, but it can be used to reduce $\KHS^g$ into a direct sum of energy eigenspaces $\KHS^g_E$
\begin{equation}
\KHS^g = \bigotimes_{\sigma^g(E)} \KHS^g_E,
\end{equation}
where $\sigma^g(E)$ is the spectrum of $\hat{H}$ for a particular $g$.

Because $\mathrm{G}_N$ is a symmetry of $\hat{H}$, the space $\KHS^g$ can also be reduced into a direct sum of
irreps of $\mathrm{G}_N$ labeled by the triplet $\mu=\{\nu_R, \pi, \pp{p} \}$, i.e.\
\begin{equation}
\KHS^g = \bigotimes_\mu \bigotimes_{\tau_\mu = 0}^\infty \KHS^g_{\mu \tau_\mu},
\end{equation}
where $\tau_\mu$ labels the degeneracy of $\mu$-type irreps. The three labels $\{\nu_R,\pi, \pp{p}\}$ of the irreps correspond to the three factors of $\mathrm{G}_N$. First, the center-of-mass excitation quantum number $\nu_R\in\mathbb{N}=\{0,1,2,\ldots\}$ labels the irreps for $\mathrm{U}(1)$. Second, the parity quantum number $\Pi=\pm 1$ labels the irrep for parity inversion. The relative parity quantum number $\pi=\Pi(-1)^{\nu_R}$ will be used instead of $\Pi$ to remove the contribution due to the parity from the center-of-mass excitations since inversion of the just the center-of-mass coordinate is contained with the $\mathrm{U}(1)$ symmetry factor.

The third factor of $\mathrm{G}_N$ is the symmetric group $\mathrm{S}_N$. Here I briefly summarize some notation for $\mathrm{S}_N$ and its irreps as necessary for this article; consult a standard reference like \cite{hamermesh_group_1989} for details. The group $\mathrm{S}_N$ has $N!$ elements corresponding to the permutations of $N$ objects. The set of elements of $\mathrm{S}_N$ decomposed into conjugacy classes for each partition $\lfloor p \rfloor$ of $N$, e.g.\ for four particles there are five partitions $\pp{4}$, $\pp{31}$, $\pp{22}\equiv\pp{2^2}$, $\pp{211}\equiv\pp{21^2}$ and $\pp{1111}\equiv\pp{1^4}$. For each conjugacy class there is also an irrep of $\mathrm{S}_N$, and these are often depicted as standard Young diagrams with $N$ boxes, e.g.\ $\pp{31}=\mbox{\tiny \yng(3,1)}$. The dimension of each irrep $\Delta\pp{p}$ is the number of standard Young tableaux for that diagram, i.e.\ the diagram's boxes are labeled such that rows of boxes always have an equal or increasing number as one moves to the right, and vertical boxes have always have increasing numbers. For a given $N$, the irrep $\pp{N}$ is one-dimensional and totally symmetric and the irrep $\pp{1^N}$ is one-dimensional and totally antisymmetric. All other irreps are multidimensional. Conjugate representations, in which rows and columns are exchanged, have the same dimension. Within an irrep, each element of $\mathrm{S}_N$ is represented by a $\Delta\pp{p}$-dimensional matrix. The traces of these matrices are called the characters of the representation, and all elements of $\mathrm{S}_N$ from the same conjugacy class have the same character.

Assuming there are no other symmetries of the Hamiltonian than $\mathrm{G}_N$ for the parameter range $0 < g < \infty$, then the degeneracy of any eigenvalue $E$ is constrained to be the dimension of an irrep of $\mathrm{G}_N$ and there is a unique correspondence between each energy degeneracy subspace $\KHS^g_E$ and a specific irrep $\KHS^g_{\mu \tau_\mu}$. This correspondence  will not change as $g$ is adiabatically tuned within the range $0 < g < \infty$, even though energy of the subspace may vary. Therefore the energy levels in the subspace $\KHS^g_\mu = \bigoplus_{\tau_\mu} \KHS^g_{\mu \tau_\mu}$ do not cross or reorder as $g$ is varied from non-interacting to hard-core repulsion. If they did cross at some $g$, this would imply an additional symmetry of $\hat{H}$ for that particular value of interaction strength. To my knowledge there is no analytic or numerical evidence for such an ``accidental'' symmetry (see for example~\cite{harshman_symmetries_2012} for the three-body case), but a proof of this assumption that $\mathrm{G}_N$ exhausts the symmetries of $\hat{H}$ for the parameter range $0 < g < \infty$ would be most welcome.

If there are no spin degrees of freedom, or only one spin component is accessible, the spin Hilbert space $\SHS$ is one-dimensional and the decomposition of $\KHS^g$ into $\mathrm{G}_N$ irreps is the complete reduction of the problem. In that case, only the irreps  $\pp{N}$ for bosons and $\pp{1^N}$ for fermions are physically realizable. If the particles have multiple accessible spin components, then there are several different ways to proceed. One way is to reduce the spin Hilbert space $\SHS$ with respect to $\mathrm{S}_N$ and then takes its direct product with the spatial Hilbert space $\KHS^g$. Note that the total spin operator $\hat{\bf S}$ is invariant with respect to $\mathrm{S}_N$ so subspaces with the same total spin and spin component will belong to an irrep of $\mathrm{S}_N$. Techniques for the reduction of a spin system are well-known~\cite{hamermesh_group_1989}; an example with four spin-$1/2$ fermions is provided in Table \ref{spintable}. Then the total Hilbert space $\mathcal{H}=\SHS \otimes \KHS^g$ can be symmetrized to account for fermions or bosons by taking the direct product of the $\mathrm{S}_N$ irreps for the spin with the $\mathrm{S}_N$ factor of the $\mathrm{G}_N$ irrep for the spatial part. The reduction of the direct product of two  $\mathrm{S}_N$ irreps will contain a single copy of $\pp{N}$ if the irreps are the same and a single copy of $\pp{1^N}$ if they are conjugate irreps. This method is essentially a variation of the technique described in \cite{guan_exact_2009} and extended in \cite{cui_ground-state_2014}.

\begin{table}
\centering
\begin{tabular}{|c|c|c|c|c|}
\hline
& $s=2$ & $s=1$ & $s=0$ & Total \\
\hline
$s_z=2$  &{\scriptsize \young(\uparrow\uparrow\uparrow\uparrow)} &  & &1\\ \hline
$s_z=1$  & {\scriptsize\young(\uparrow\uparrow\uparrow\downarrow)} & {\scriptsize\young(\uparrow\uparrow\uparrow,\downarrow)} & & 4\\\hline
$s_z=0$  & {\scriptsize\young(\uparrow\uparrow\downarrow\downarrow)} & {\scriptsize\young(\uparrow\uparrow\downarrow,\downarrow)} & {\scriptsize\young(\uparrow\uparrow,\downarrow\downarrow)} & 6 \\\hline
$s_z=-1$  & {\scriptsize\young(\uparrow\downarrow\downarrow\downarrow)} & {\scriptsize\young(\uparrow\downarrow\downarrow,\downarrow)} & & 4 \\\hline
$s_z=-2$  & {\scriptsize\young(\downarrow\downarrow\downarrow\downarrow)} &  & & 1\\
\hline
Total & $ D^2 $ & $3 D^2$  & $2 D^0$ & \\ \hline
\end{tabular}
\caption{This table summarizes the decomposition of the spin Hilbert space $\SHS\sim \mathcal{C}^{16}$ for $N=4$ spin-$1/2$ particles into irreps of $\mathrm{S}_4$. For example, this shows that there are three ways to construct a spin state with $s=1$ and $s_z=1$, and they transform among themselves under particle permutations as the irrep $\pp{31}$. In total, $\SHS$ can either be decomposed as $D^2 \oplus 3 D^1 \oplus 2 D^0$, where $D^s$ is the irrep of $\mathrm{SU}(2)$, or as $ 5 \pp{4} \oplus 3 \pp{31} \oplus \pp{2}$.}
\label{spintable}
\end{table}

Another equivalent method, which is applied in the remainder of this article, is to work only with the spatial states and treat particles with different spin components as if they were distinguishable (c.f. \cite{gharashi_correlations_2013}). Then although $\hat{H}$ still has the full $\mathrm{S}_N$ symmetry, the physically realizable states have a symmetry that is a subgroup of $\mathrm{S}_N$. For example, consider four two-component fermions with two particles spin-up, two spin-down. To account for symmetrization, the only possible states are in those irreps of $\mathrm{S}_4$ which when reduced by $\mathrm{S}_2\times \mathrm{S}_2$ contain the irrep $\pp{1^2}\times\pp{1^2}$. As an example, see Table \ref{tab:4red} for the reduction of $\mathrm{S}_4$ with respect to all bosonic and fermionic irreps of subgroup. See Appendix A for similar tables for $N=3$ and $N=5$. The advantage of this second method is that the symmetrized spatial states can be constructed without explicitly constructing the spin state.

\begin{table}
\centering
\begin{tabular}{|c|c|ccccc|}
\hline
number of & component & \multicolumn{4}{c}{ $\pp{p}$} &\\
components &pattern &  \pp{4} & \pp{31} & \pp{2^2} & \pp{21^2} & \pp{1^4} \\
\hline
1 &$(4)_B$ &     1 & 0 & 0 & 0 & 0 \\
2 & $(31)_B$  &  1 & 1 & 0 & 0 & 0 \\
2 & $(22)_B$  &  1 & 1 & 1 & 0 & 0 \\
3 & $(211)_B$ &  1 & 2 & 2 & 1 & 0 \\ \hline
1 & $(4)_F$   &  0 & 0 & 0 & 0 & 1 \\
2 & $(31)_F$  &  0 & 0 & 0 & 1 & 1 \\
2 & $(22)_F$   & 0 & 0 & 1 & 1 & 1 \\
3 & $(211)_F$  & 0 & 1 & 2 & 2 & 1 \\ \hline
4 & $(1111)$   & 1 & 3 & 2 & 3 & 1 \\
\hline
\end{tabular}

\caption{This table gives the reduction of $\mathrm{S}_4$ irreps by subgroups to account for multi-component bosons and fermions. The notation $(22)_B$ means two bosons in one component, and two bosons in another component. The case $(1111)$ corresponds to distinguishable particles.}
\label{tab:4red}
\end{table}

\section{Non-interacting solutions}

At $g=0$, the Hamiltonian (\ref{hammy}) is equivalent to an $N$-dimensional isotropic harmonic oscillator with $\mathrm{U}(N)$ symmetry~\cite{jauch_problem_1940, baker_degeneracy_1956, louck_group_1965}. Energy subspaces $\KHS^0_X \equiv \KHS^0_E$ have energy $E= \hbar\omega(X+N/2)$ with $X\in\mathbb{N}$ and degeneracy
\begin{equation}\label{dim}
d^X_N = \frac{(X+N-1)!}{X!(N-1)!}.
\end{equation}
A standard set of invariants to diagonalize the degeneracy of $\KHS_X^0$ are the $N$ quantum numbers $\{n_1,n_2,\ldots,n_N\}$ representing the excitation number for each particle treated as separable harmonic oscillators. However, the reduction into irreps takes the simplest form when the $N$-dimensional harmonic oscillator is solved in the Jacobi hypercylindrical coordinates $\{R, \rho, \hat{\Omega}\}$. The coordinate
\[
R = \frac{1}{\sqrt{N}}\sum_{i=1} x_i
\]
is the normalized center-of-mass. The hypercylindrical radius $\rho$ is the distance from the $R$-axis
\[
\rho =  \frac{1}{\sqrt{N}}  \left((N-1) \sum_{i=1}  x_i - 2 \sum_{i<j}x_i x_j\right)^{1/2}
\]
and the other relative coordinates are the $(N-2)$ hyperangles $\hat{\Omega}$. Note that the definitions of $R$ and $\rho$ are $\mathrm{S}_N$ invariant.

Quantization in this coordinate system leads to the alternate basis $\kt{\nu_R,\nu_\rho,\lambda,\{\chi\}}$, where  $\nu_R\in\mathbb{N}$ is the excitation quantum number for $R$ and $\nu_\rho\in\mathbb{N}$ is the excitation quantum number associated for $\rho$. The quantum number $\lambda$ is related to the eigenvalues of the grand angular momentum operator $\hat{\Lambda}$ for the hyperangular coordinates $\hat{\Omega}$~\cite{yanez_position_1994}. For $N>3$ the degeneracy of a $\lambda$ subspace is (modifying Eq.~(3.86) from \cite{avery_hyperspherical_2000})
\[
\epsilon_\lambda^N = \frac{(N + 2 \lambda -3)(\lambda + N -4)!}{\lambda! (N-3)!}
\]
and for $N=3$, $\epsilon^3_{\lambda=0}= 1$ and $\epsilon^3_{\lambda\neq 0}= 2$. The numbers $\{\chi\}$ are the set of parameters that label an orthonormal basis for the $\lambda$ subspace. This degeneracy can be diagonalized many ways (e.g.~\cite{yanez_position_1994, aquilanti_hyperspherical_1998}), but the particular choice is not important for the results of this article.

Note that the quantum numbers $\{\nu_R,\nu_\rho,\lambda\}$ are invariant under $\mathrm{G}_N$, and two of the irrep labels can be immediately determined: $\nu_R$ itself and $\pi=(-1)^\lambda$. The energy expressed in the hypercylindrical quantum numbers is $\hbar\omega(\nu_R + 2 \nu_\rho + \lambda + N/2)$. Reducing $\KHS^0_X$ into subspaces $\KHS^0_{\nu_R,\nu_\rho,\lambda}$ with $X =\nu_R+ 2\nu_\rho+ \lambda$  is a straightforward arithmetic problem that is independent of $N$. Each subspace $\KHS^0_{\nu_R,\nu_\rho,\lambda}$ must then be further reduced into $\mathrm{S}_N$ irreps $\pp{p}$. The key observation is that which irreps of $S_N$ appear depends only on the grand angular momentum $\lambda$. 

Here is an outline of the $\lambda$ subspace reduction method: Assume that the reduction has already been found for all values of $\lambda$ up to $X-1$. To find the reduction for $\lambda = X$, first reduce $\KHS^0_X$ into irreps of $\mathrm{S}_N$ using the symmetries of the partitions of $X$. For example, when $N=4$ and $X=2$ the partitions are $(2,0,0,0)$ and $(1,1,0,0)$. The first partition corresponds to a four-dimensional particle excitation basis that can be reduced into $\pp{4}\oplus\pp{31}$, the second to a six-dimensional basis that reduces to $\pp{4}\oplus\pp{31}\oplus\pp{2^2}$.  Second, decompose $\KHS^0_X$ into $\KHS^0_{\nu_R,\nu_\rho,\lambda}$ subspaces. Continuing with the example, $\KHS^0_2 = \KHS^0_{2,0,0}\oplus\KHS^0_{1,0,1}\oplus\KHS^0_{0,1,0}\oplus\KHS^0_{0,0,2}$. One of these will always be the subspace $\KHS_{0,0,X}$ and the rest have $\lambda$'s for which the $\mathrm{S}_N$ reduction problem is already solved. Subtracting the irreps for the values of $\lambda<X$, the only remaining irreps will correspond to $\lambda=X$. See the results of the reduction for $N=4$ in Table \ref{tab:gzero4}.

See Appendix B for a more detailed description of how the $\lambda$ reduction is calculated, and for tables summarizing the the $\lambda$ reduction of $\KHS^0_{\nu_R,\nu_\rho,\lambda}$ for $N=3$ and $N=5$. For all $N$, when $\lambda =0 $, the space $\KHS^0_{\nu_R,\nu_\rho,0}$ carries only the single irrep $\pp{N}$ and when $\lambda =1 $, the space $\KHS^0_{\nu_R,\nu_\rho,1}$ carries only the irrep $\pp{N\!-\!1 \,1}$. For $N \geq 4$, the space  $\KHS^0_{\nu_R,\nu_\rho,\lambda}$ will not be simply reducible for general $\lambda$.

\begin{table}
\centering
\begin{tabular}{|r|c|c|c|c|c||r|c|c|c|c|c|}
\hline
$\lambda$ & $\pp{4}$ & $\pp{31}$ & $\pp{2^2}$ & $\pp{21^2}$ & $\pp{1^4}$ & $\lambda$ & $\pp{4}$ & $\pp{31}$ & $\pp{2^2}$ & $\pp{21^2}$ & $\pp{1^4}$ \\
\hline
0  & 1 & 0 & 0 & 0 & 0 & 7 & 1 & 2 & 1 & 2 & 0\\
 1 & 0 & 1 & 0 & 0 & 0 &8  & 1 & 2 & 2 & 2 & 0 \\
2  & 0 & 1 & 1 & 0 & 0 & 9 & 1 & 3 & 1 & 3 & 1\\
   3 & 1 & 1 & 0 & 1 & 0 &10 & 1 & 3 & 2 & 2 & 1 \\
4  & 1 & 1 & 1 & 1 & 0 &  11 & 1 & 3 & 2 & 3 & 0\\
  5 & 0 & 2 & 1 & 1 & 0 &12 & 2 & 3 & 2 & 3 & 1 \\
6  & 1 & 2 & 1 & 1 & 1 & 13 & 1 & 4 & 2 & 3 & 1\\
\hline
\end{tabular}
\caption{This table gives the degeneracy of $\mathrm{S}_4$ irreps in each $\KHS^0_{\nu_R,\nu_\rho,\lambda}$ subspace. To calculate for $\lambda = 12 + \lambda'$  add the irrep degeneracy pattern $\{1,3,2,3,1\}$ to the pattern for $\lambda'$.}
\label{tab:gzero4}
\end{table}

There are alternate methods to achieve this reduction. For the case of $N=3$, this reduction was done by the author in \cite{harshman_symmetries_2012} using explicit construction of representations of $\mathrm{S}_3\times \mathrm{Z}_2$ on Jacobi hypercylindrical basis functions. For the case of $N=4$, the reduction can be performed by recognizing that all permutations in $\mathrm{S}_4\times \mathrm{Z}_2$ is isomorphic to $\mathrm{O}_h$, the point symmetry group of a cube. Using well-known properties of spherical harmonics, the characters of the representation of these operations on $\lambda$ subspaces can be calculated and then reduced by characters of the irreps of $\mathrm{S}_4\times \mathrm{Z}_2$ using standard group representation techniques. These reproduce known results for tetrahedral harmonics from molecular 
physics, c.f.~\cite{wormer_tetrahedral_2001}.

However the $\lambda$ reduction is accomplished, it completes the decomposition of $\KHS^0$ into irreps of $\mathrm{G}_N$:
\begin{eqnarray}
\KHS^0 &=& \bigotimes_{X=0}^\infty \KHS^0_X \nonumber \\
\KHS^0_X & =&  \bigotimes_{\nu_R + 2 \nu_\rho +\lambda = X} \KHS^0_{\nu_R,\nu_\rho,\lambda}\nonumber\\
\KHS^0_{\nu_R,\nu_\rho,\lambda} &= &\bigotimes_{\pp{p}} \bigotimes_{\tau_{\pp{p}}} \KHS^0_{\nu_R,\nu_\rho,\lambda; \pp{p} \tau_{\pp{p}}} 
\end{eqnarray}
Energy eigenfuctions for $g=0$ in position space ${\bf x}=(x_1,\ldots,x_N)$ can be uniquely labeled as
\[
\bk{{\bf x}}{\nu_R,\nu_\rho,\lambda; \pp{p}, \tau; j},
\]
where $\tau$ labels the degeneracy of $\pp{p}$ in the $\KHS^0_{\nu_R,\nu_\rho,\lambda}$ subspace and $j$ labels the $\Delta\pp{p}$-fold degeneracy within the $\pp{p}$ irrep.

For single-component bosons or fermions, the spin wave function can only be symmetric. Therefore, as long as irrep $\pp{N}$ (or $\pp{1^N}$) appears in the $\mathrm{S}_N$ reduction of $\lambda$, then the wave function $\bk{{\bf x}}{\nu_R,\nu_\rho,\lambda; \pp{N} \tau}$ (or $\bk{{\bf x}}{\nu_R,\nu_\rho,\lambda; \pp{1^N} \tau}$)  has the right symmetry to represent single-component bosons (or fermions). Multi-component fermionic or bosonic symmetrization can be incorporated using the direct product of decomposition of $\mathrm{S}_N$ or the subgroup reduction method described in the previous section. For example, states representing four two-component fermions, three spin-up and one spin-down, have the symmetry subgroup $\mathrm{S}_3\times \mathrm{S}_1\equiv \mathrm{S}_3$. The $\mathrm{S}_3$ subgroup irrep that antisymmetrizes the three spin-up particles is $\pp{1^3}$. The $\mathrm{S}_4$ irrep $\pp{21^2}$ reduces with respect to $\mathrm{S}_3$ irreps according to $\pp{21^2} = \pp{21} \oplus \pp{1^3}$. That means there is one state within each $\pp{21^2}$-type subspace that can represent a three spin-up, one spin-down fermionic state, and such a state is labeled unambiguously as $\kt{\nu_R,\nu_\rho,\lambda; \pp{21^2} \tau; \pp{1^3}}$.

\begin{table}
\centering
\begin{tabular}{|c|ccccccccccccc|}
\hline
component &  \multicolumn{13}{c|}{$\lambda$}\\
pattern &  0 & 1& 2& 3& 4 & 5 & 6 & 7 & 8 & 9 & 10 & 11 & 12 \\
\hline
$(4)_B$ &     1 & 0 & 0 & 1 & 1 & 0 & 1 & 1 & 1 & 1  & 1 & 1 & 2\\
$(31)_B$ &  1 & 1 & 1 & 2 & 2 & 2 & 3 & 3 & 3 & 4  & 4 & 4 & 5\\
$(22)_B$ &  1 & 1 & 2 & 2 & 3 & 3 & 4 & 4 & 5 & 5  & 6 & 6 & 7\\
$(211)_B$ &  1 & 2 & 3 & 4 & 5 & 6 & 7 & 8 & 9 & 10 & 11 & 12 & 13\\ \hline
$(4)_F$ &  0 & 0 & 0 & 0 & 0 & 0 & 1 & 0 & 0 & 1  & 1 & 0 & 1\\
$(31)_F$ &  0 & 0 & 0 & 1 & 1 & 1 & 2 & 3 & 3 & 3  & 3 & 3 & 4\\
$(22)_F$ & 0 & 0 & 1 & 1 & 2 & 2 & 3 & 3 & 4 & 4  & 5 & 5 & 6\\
$(211)_F$ & 0  & 1 & 2 & 3 & 4 & 5 & 6 & 7 & 8 & 9 & 10 & 11 & 12 \\ \hline
$(1111)$ & 1 & 3 & 5 & 7 & 9 & 11 & 13 & 15 & 17 & 19 & 21 & 23 & 25\\
\hline
\end{tabular}
\caption{This table incorporates the results of Tables \ref{tab:4red} and \ref{tab:gzero4} to enumerate the degeneracy multi-component states obeying symmetrization rules in the space $\KHS^0_{\nu_R,\nu_\rho,\lambda}$ for $N=4$. Note that the degeneracy of $(1111)$ is $\epsilon^4_\lambda = (2\lambda +1)$.}
\label{tab:4lam}
\end{table}

\section{Infinite Repulsion Limit}

Configuration space can be divided into $N!$ sectors, each labeled by a permutation $\langle p \rangle = \langle p_1 p_2 \ldots p_N \rangle$ of $\{12\ldots N\}$. For example with $N=4$, $\langle 1234 \rangle$ labels the sector with $x_1>x_2>x_3>x_4$. Consider any totally antisymmetric $g=0$ wave function $\bk{{\bf x}}{\nu_R,\nu_\rho,\lambda; \pp{1^N}}$. This function will have nodes at the boundaries of the sectors and from it one can construct a set of $N!$ orthonormal eigenfunctions of $\hat{H}$ in the limit $g\rightarrow \infty$, each with support on different sectors~\cite{deuretzbacher_exact_2008}:
\begin{equation}
\cbk{{\bf x}}{\nu_R,\nu_\rho,\lambda; \langle p \rangle } = \left\{ \begin{array}{ll} \sqrt{N!} |\bk{{\bf x}}{\nu_R,\nu_\rho,\lambda; [1^N]}| & {\bf x} \in \langle p \rangle \\ 0 & {\bf x} \notin \langle p \rangle \end{array} \right.
\end{equation}
These form a ``snippet'' basis~\cite{fang_exact_2011} for the subspaces $\KHS^\infty_{\nu_R,\nu_\rho,\lambda}$ which have energy $\hbar\omega(\nu_R + 2 \nu_\rho + \lambda + N/2)$. For $N=3$, there is a single fermionic state to use as a snippet base whenever $\lambda=3,6,9,\ldots, 3k, \ldots$. For $N=4$, the first few subspaces in which a single $\pp{1^4}$ irrep appears are $\lambda = 6,9,10,12,13$, and for $N=5$ the first few are $\lambda = 10, 13, 14,15$.

Reducing the subspaces $\KHS^\infty_{\nu_R,\nu_\rho,\lambda}$ into irreps of $\mathrm{G}_N$ is accomplished in the following steps. First, notice that the $2 (N!)$ elements $\gamma\in\mathrm{S}_N\times Z_2$ are represented on the $N!$-dimensional basis $\ckt{\nu_R,\nu_\rho,\lambda; \langle p \rangle }$ in a natural way. The representation $\hat{U}(c)$ of an element $c \in \mathrm{S}_N$ permutes $\langle p \rangle$ into a new order
\begin{equation}
\hat{U}(c) \ckt{\nu_R,\nu_\rho,\lambda; \langle p \rangle } = \ckt{\nu_R,\nu_\rho,\lambda; \langle p' \rangle }
\end{equation}
where $\langle p' \rangle$ is the rearrangement of $p$ corresponding to the permutation $c$. Parity inversion $\hat{U}_\Pi$ reverses the order and pulls out a factor of the intrinsic parity
\begin{equation}
\hat{U}_\Pi \ckt{\nu_R,\nu_\rho,\lambda; \langle p \rangle } = (-1)^{\nu_R + \lambda} \ckt{\nu_R,\nu_\rho,\lambda; \langle \tilde{p} \rangle }
\end{equation}
where $ \langle \tilde{p} \rangle = \langle p_N p_{N-1 \ldots p_1} \rangle$. Once the characters $\chi(\gamma)$ of the representation $\hat{U}(\gamma)$ are found from these definitions (see Appendix C for more details), then the reduction into irreps $\pp{p}^\pi$ of $\mathrm{S}_N\otimes Z_2$  is performed using the standard formula
\begin{equation}
a_\mu = \frac{1}{2 N!} \sum_\gamma \chi^{\mu *} (\gamma) \chi(\gamma)
\end{equation}
where $a_\mu$ is the number of times the irrep $\mu$ appears in the compound representations with characters $\chi$ and $\chi^{\mu *}(\gamma)$ is the character of the representation of $\gamma$ in the irrep $\mu$~\cite{hamermesh_group_1989}.

The snippet basis vectors can be projected into irreps of $\mathrm{S}_N\times \mathrm{Z}_2$ within  $\KHS^\infty_{\nu_R,\nu_\rho,\lambda}$ using the characters and standard techniques~\cite{hamermesh_group_1989}. These states are denoted $\ckt{\nu_R,\nu_\rho,\lambda; \pp{p}^\pi, \tau; j}$. As before, $\tau$ labels the degeneracy of $\pp{p}^\pi$  in the $\KHS^\infty_{\nu_R,\nu_\rho,\lambda}$ subspace and $j$ labels the degeneracy within the irrep $\pp{p}^\pi$. For $N=4$, results of this reduction are in Table \ref{tab:ginfty4}. See Appendix C for details about characters and reductions for $N=3$ and $N=5$.

\begin{table}
\centering
\begin{tabular}{|c|c|c||c|c|c|}
\hline
irrep    & even  &  odd & irrep    & even  &  odd \\
$\pp{p}^+$ &  $\lambda$ & $\lambda$ & $\pp{p}^+$ &  $\lambda$ & $\lambda$\\
\hline
$\pp{4}^+$ & 1 & 0 & $\pp{4}^-$ & 0 & 1 \\
$\pp{31}^+$ & 1 & 2 & $\pp{31}^-$ & 2 & 1 \\
$\pp{2^2}^+$ & 2 & 0 & $\pp{2^2}^-$ & 0 & 2 \\
$\pp{21^2}^+$ & 1 & 2 & $\pp{21^2}^-$ & 2 & 1 \\
$\pp{1^4}^+$ & 1 & 0 & $\pp{1^4}^-$ & 0 & 1 \\
\hline
\end{tabular}
\caption{This table gives the reduction of the $24$ dimensional subspace $\KHS^\infty_{\nu_R,\nu_\rho,\lambda}$ into irreps of $\mathrm{S}_4\times \mathrm{Z}_2$ labeled as $\pp{p}^\pi$.}
\label{tab:ginfty4}
\end{table}

\section{Mapping Examples and Conclusions}

To conclude, I present a few examples of states and mappings.
For $N=3$, there are six possible irreps of $\mathrm{S}_3\times\mathrm{Z}_N$. The subspaces $\KHS^0_{\nu_R,\nu_\rho,\lambda}$ and $\KHS^\infty_{\nu_R,\nu_\rho,\lambda}$ are simply reducible, i.e.\ each irrep $\pp{p}^\pi$ only appears once for a given $\{\nu_R,\nu_\rho,\lambda\}$ and there are no $\tau$ labels required.  The antisymmetric states $\kt{\nu_R,\nu_\rho,\lambda; \pp{1^3}}=\ckt{\nu_R,\nu_\rho,\lambda; \pp{1^3}}$ with $\lambda = 3,6,9,\ldots$ are eigenstates for all values of $g$, and the symmetric states $\kt{\nu_R,\nu_\rho,\lambda; \pp{3}}$ with $\lambda = 0,3,6,\ldots$ map to  $\ckt{\nu_R,\nu_\rho,\lambda+3; \pp{3}^\pi}$ with $\pi = (-1)^\lambda$ (i.e.~the corresponding Girardeau states). The states $\kt{\nu_R,\nu_\rho,\lambda; \pp{21}; j}$ ($j=1,2$) can be reduced by the subgroup $\mathrm{S}_2$ into a two-component bosonic state $\kt{\nu_R,\nu_\rho,\lambda; \pp{21}; \pp{2}}$ and a two-component fermionic state $\kt{\nu_R,\nu_\rho,\lambda; \pp{21}; \pp{1^2}}$. The ground state of the two-component fermion sector is $\kt{0,0,1; \pp{21}; \pp{1^2}}$, and it maps into the lowest energy state with matching irrep labels $\mu=\{0,-,\pp{21}\}$, which is $\ckt{0,0,3; \pp{21}^-}$. Explicit construction shows this is the same state constructed in \cite{gharashi_correlations_2013,volosniev_strongly-interacting_2013}. 

\begin{figure}
\centering
\includegraphics[width=0.48\linewidth]{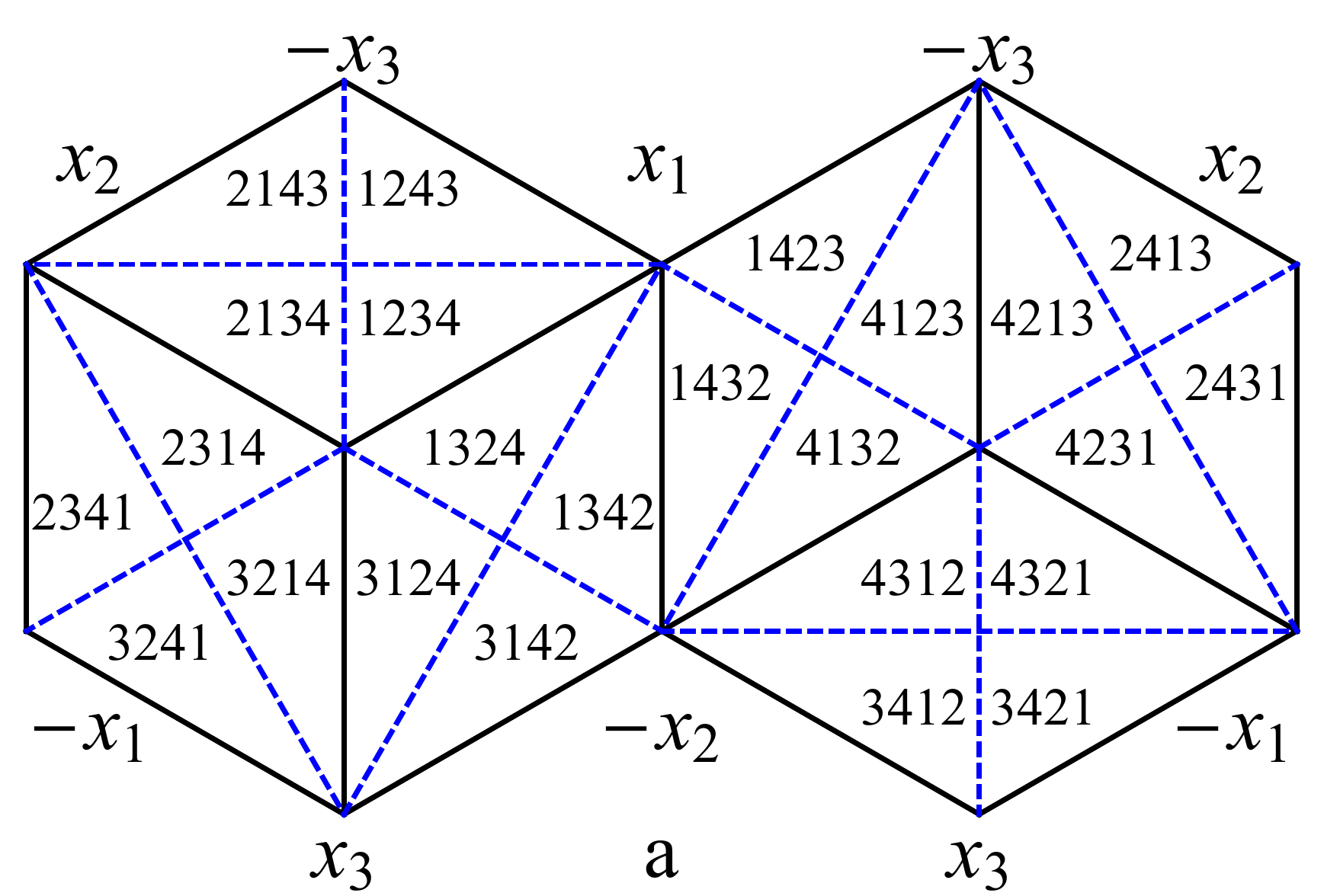}
\includegraphics[width=0.48\linewidth]{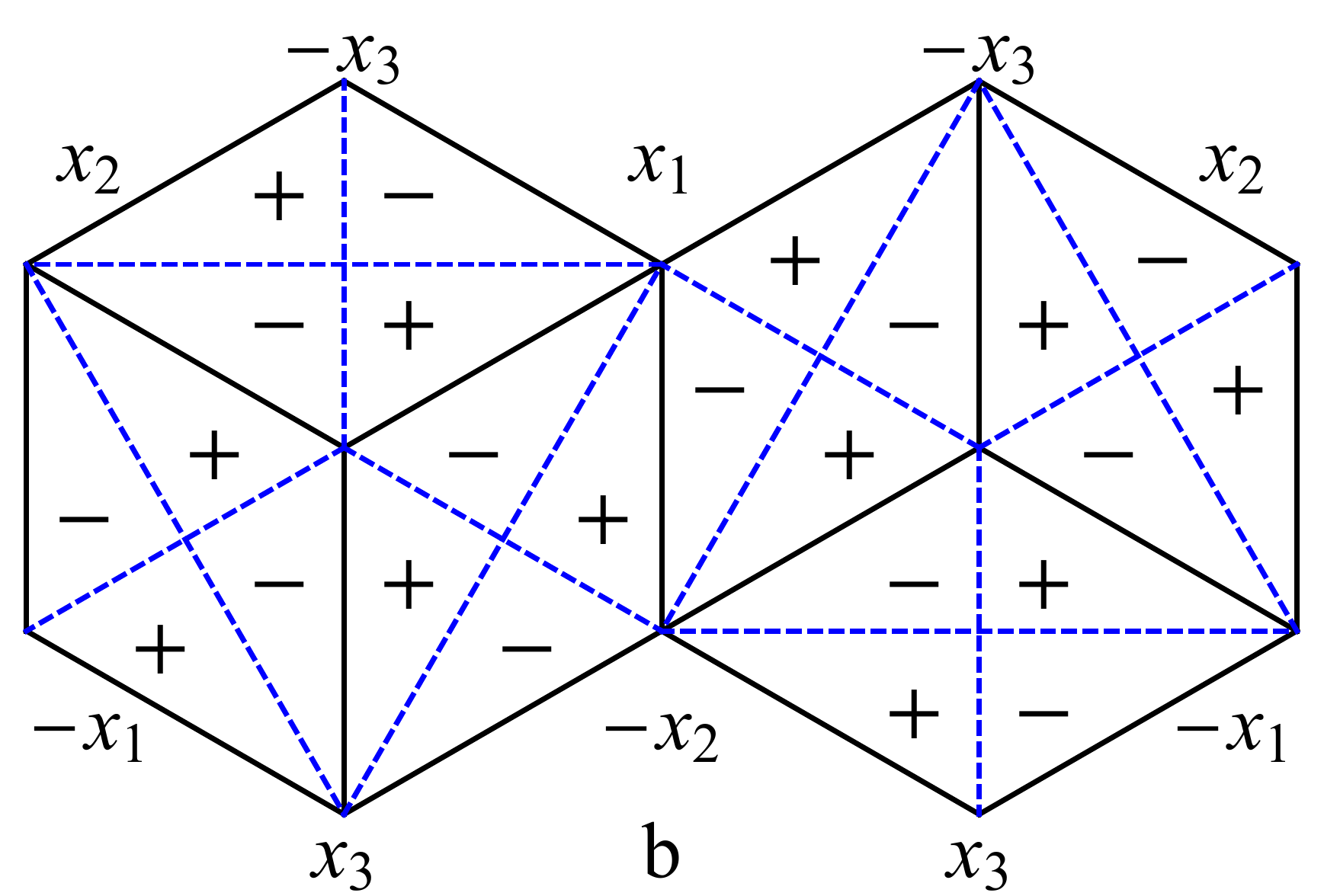}\\
\includegraphics[width=0.48\linewidth]{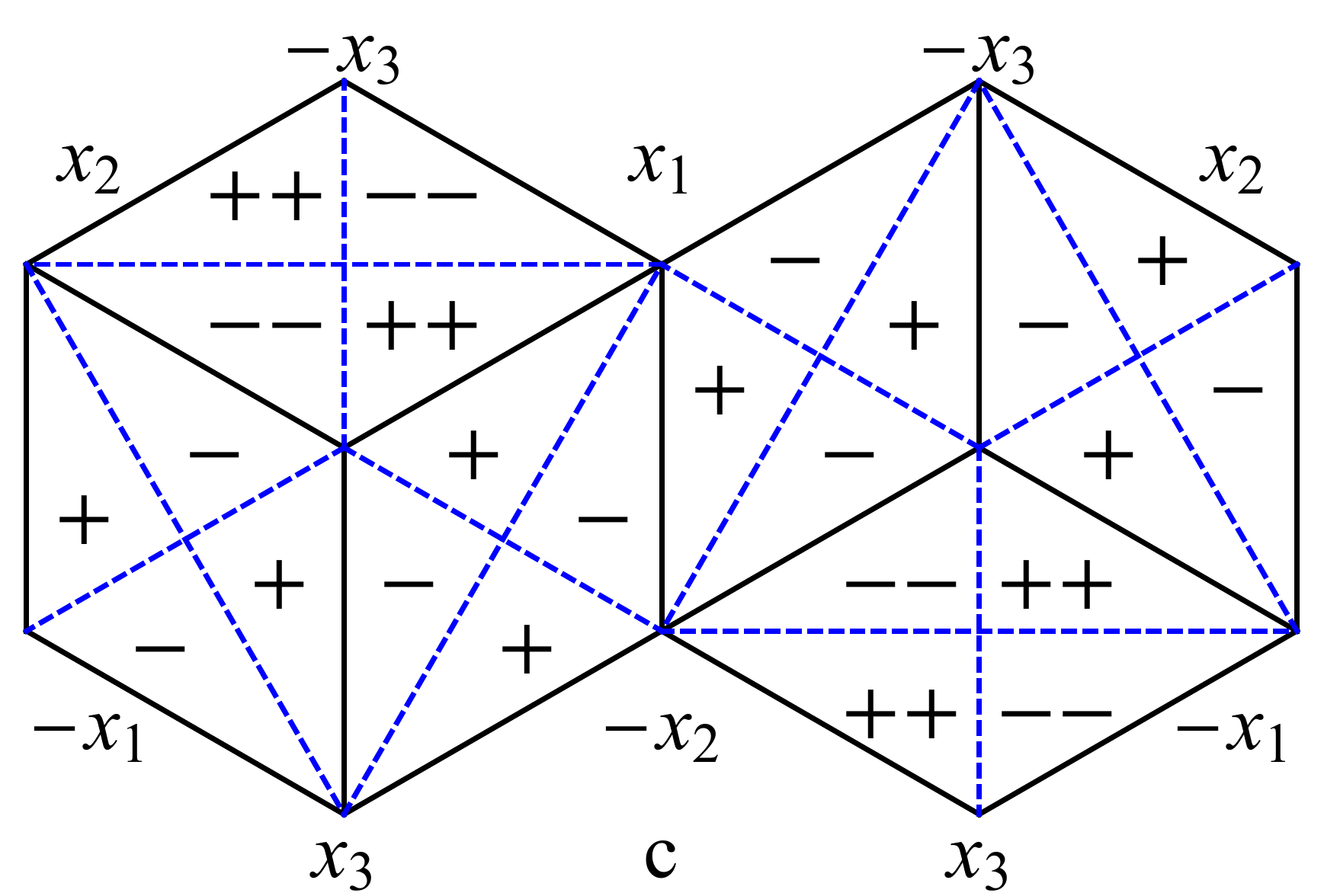}
\includegraphics[width=0.48\linewidth]{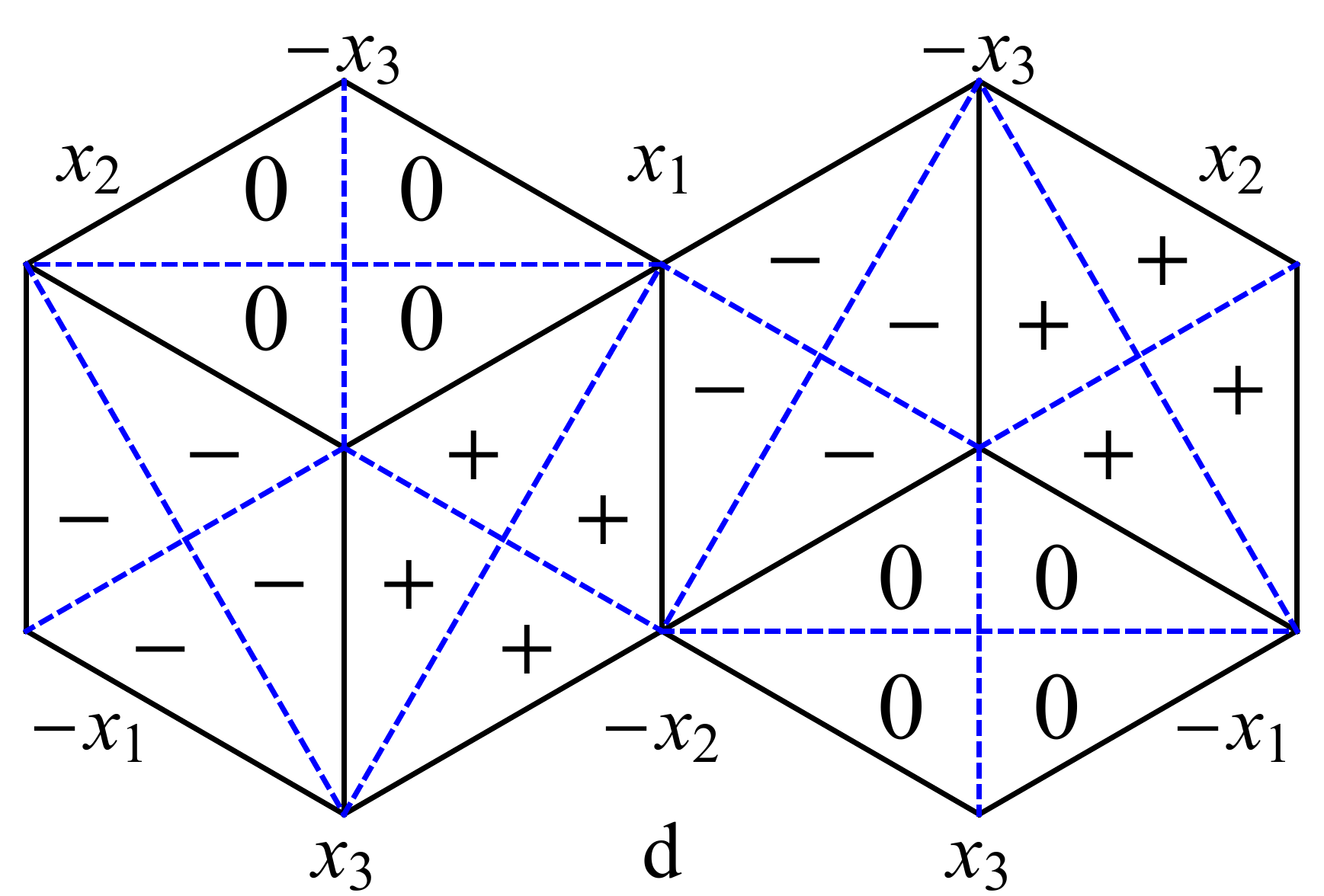}\\
\includegraphics[width=0.48\linewidth]{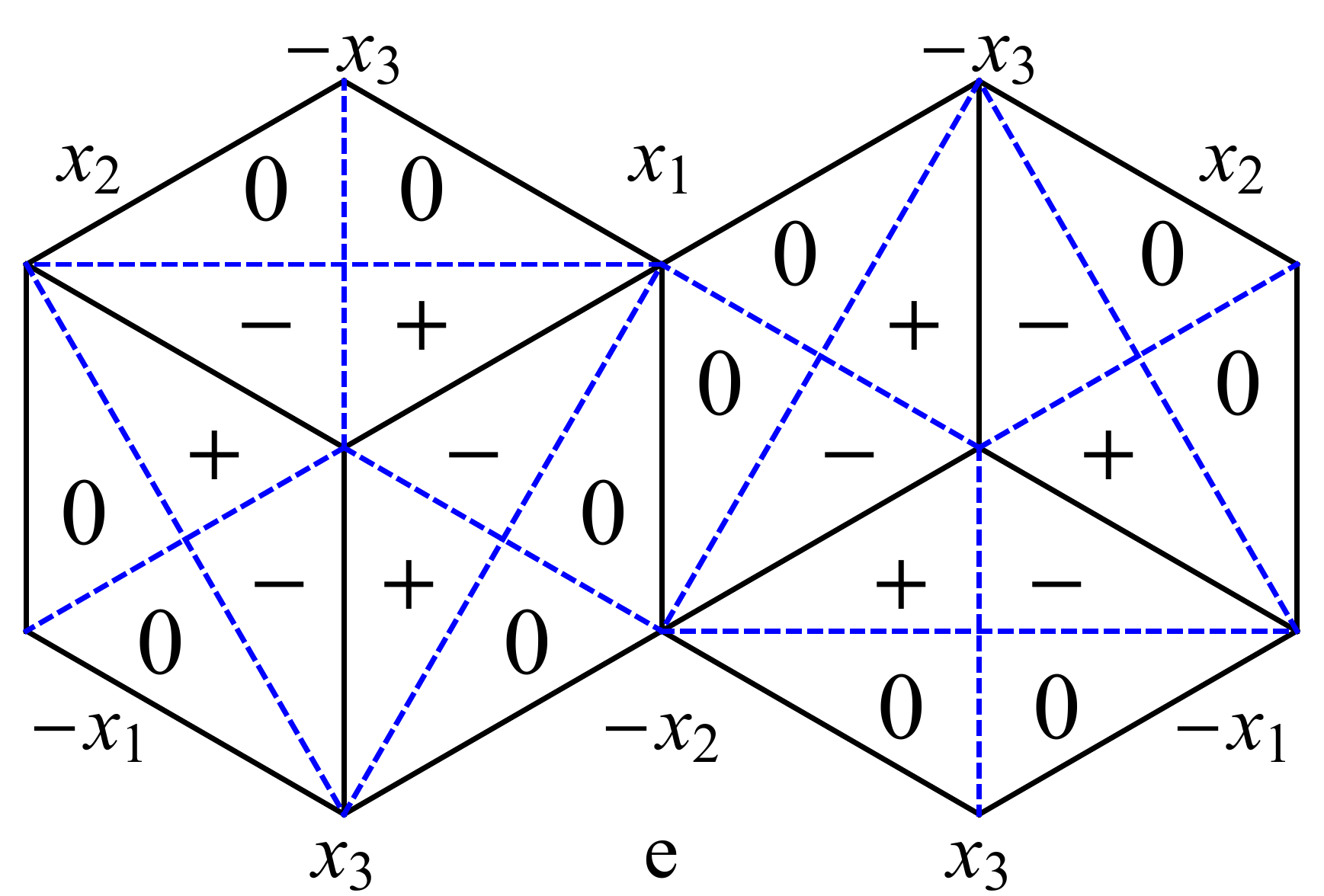}
\includegraphics[width=0.48\linewidth]{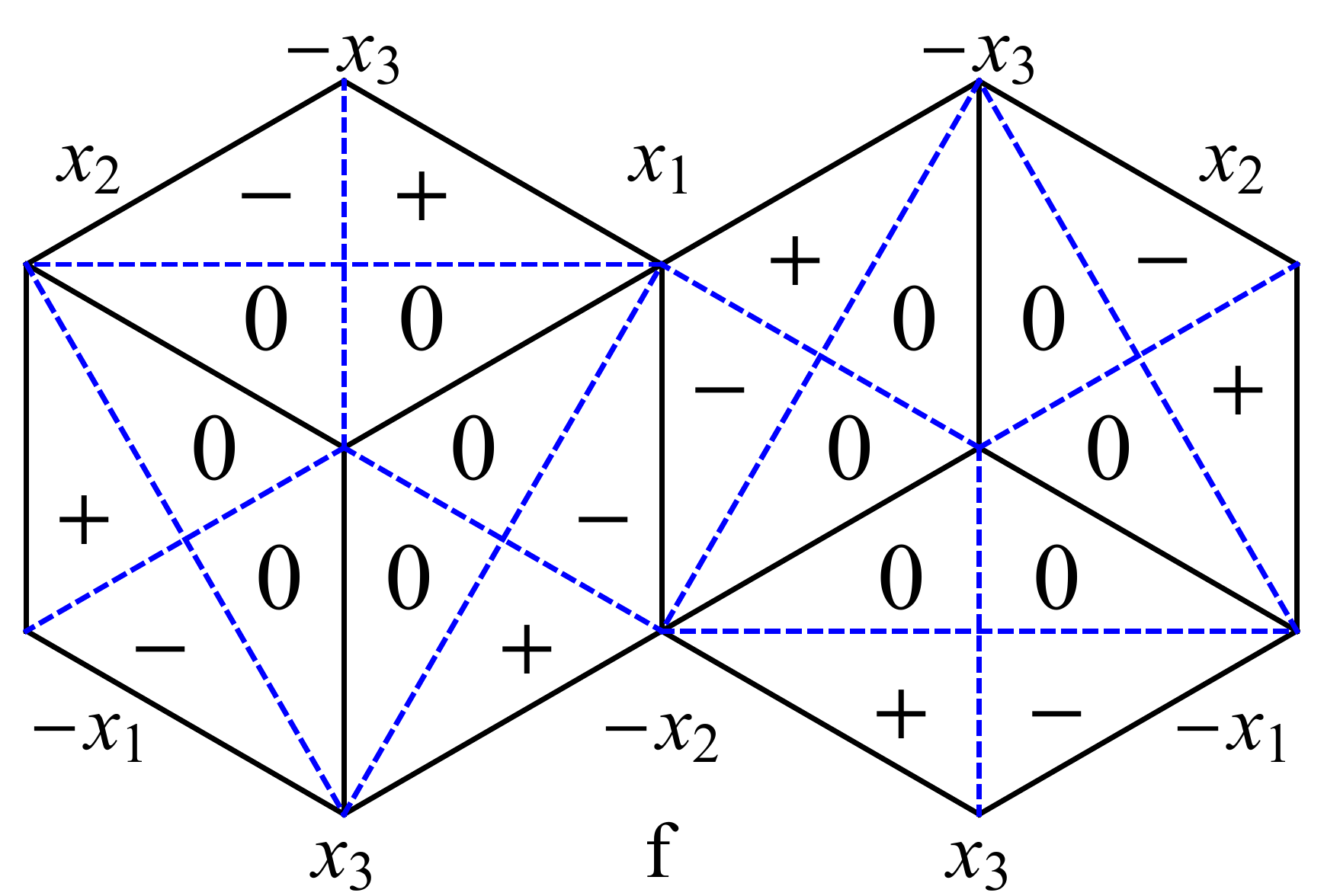}\\
\caption{(color online) These figures schematically depict the fermionic ground states of four particles in the limit $g \rightarrow \infty$. (a) Configuration space is broken into 24 sectors, each with a specific ordering of particles, e.g. $4231 \rightarrow x_4>x_2>x_3>x_1$. Each of these sectors is represented by a triangle and all wave functions must be zero on the sides of the triangle. The center-of-mass and radial components of the wave functions are not depicted in these diagrams, just the angular part. (b) A state in $\pp{1^4}^+$: The same function is pasted into each triangle, with the relative phase represented by pluses and minuses. All positive-parity, single component fermionic states, including the ground state, have this form. (c) and (d) The two states (\ref{22}) in $\pp{2^2}^+$ reduced by the subgroup $\mathrm{S}_2\times\mathrm{S}_2$ into the irrep $\pp{1^2}\times\pp{1^2}$. The ground state of the two spin-up, two spin-down configuration adiabatically maps into the space spanned by wave functions with this form. The notation $++$ and $--$ means the amplitude in that section is twice as much, and zero means no amplitude. (e) and (f) The two states (\ref{212}) in $\pp{21^2}^-$ reduced by the subgroup $\mathrm{S}_3$ into the irrep $\pp{1^3}$. The ground state of the three spin-up, one spin-down configuration adiabatically maps into the space spanned by wave functions with this form. }
\label{fig1} 
\end{figure}

For $N=4$, the situation is more complicated because the subspaces $\KHS^0_{\nu_R,\nu_\rho,\lambda}$ and $\KHS^\infty_{\nu_R,\nu_\rho,\lambda}$ are not simply reducible and because the degeneracy $\epsilon^4_\lambda$ of  $\KHS^0_{\nu_R,\nu_\rho,\lambda}$ grows as $2(\lambda +1)$ instead of staying fixed like the $N=3$ case. Figure \ref{fig1} shows a schematic representation of the $N=4$ fermionic ground state snippet solutions. The one-component fermionic ground state is $\kt{0,0,6;\pp{1^4}} = \ckt{0,0,6;\pp{1^4}^+}$. The two component fermionic ground states for $g=0$ are
\[
\kt{0,0,2; \pp{2^2}; \pp{1^2}\!\times\!\pp{1^2}}
\]
for the two spin-up, two-spin down configuration and 
\[
\kt{0,0,3; \pp{21^2},\pp{1^3}}
\]
for the three spin-up, one spin-down (or reverse) configuration. Both of these map to $\KHS^\infty_{0,0,6}$ as $g \rightarrow \infty$, but the corresponding $\mathrm{S}_4$ irreps $\pp{2^2}^+$ and $\pp{21^2}^-$ are doubly-degenerate so the exact adiabatic state must be a superposition of two basis states
\begin{eqnarray}\label{22}
&\ckt{0,0,6; \pp{2^2}^+ 1; \pp{1^2}\!\times\!\pp{1^2}}&\ \mbox{and}\nonumber\\
&\ckt{0,0,6; \pp{2^2}^+ 2; \pp{1^2}\!\times\!\pp{1^2}}&
\end{eqnarray}
for the two spin-up, two-spin down configuration and 
\begin{equation}\label{212}
\ckt{0,0,6; \pp{21^2}^- 1; \pp{1^3}}\ \mbox{and}\ \ckt{0,0,6; \pp{21^2}^+ 2; \pp{1^3}}
\end{equation}
for the three spin-up, one spin-down (or reverse) configuration. The results here are consistent with state constructions using other analytical approaches, such as \cite{cui_ground-state_2014, volosniev_strongly-interacting_2013, deuretzbacher_quantum_2013, volosniev_multicomponent_2013}, and with numerical solutions like \cite{sowinski_few_2013}.

For $N=5$, the complications brought on by the fact that the spaces $\KHS^0_{\nu_R,\nu_\rho,\lambda}$ and $\KHS^\infty_{\nu_R,\nu_\rho,\lambda}$ are not simply reducible become even larger. The two component fermionic ground states are
\[
\kt{0,0,4;\pp{2^2 1},\pp{1^3}\!\times\!\pp{1^2}}
\]
 for three up and two down (or reverse) and
 \[
 \kt{0,0,6;\pp{2 1^3},\pp{1^4}}
 \]
 for four up and one down (or reverse). These two states map adiabatically into three- and two-dimensional subspaces of $\KHS^\infty_{0,0,10;\pp{2^2 1}^+}$ and $\KHS^\infty_{0,0,10;\pp{21^3}^+}$, respectively. Figuring out how to identify which superposition of these symmetrized snippet basis vectors are appropriate for $N>3$ is the subject of ongoing work. 

In summary, I have presented a spectroscopic method based on the reduction of energy eigenspaces with respect to the symmetry group $\mathrm{G}_N$. Every state belongs to an irrep labeled by center-of-mass excitation, relative parity and permutation group symmetry. This group appears to be the maximal symmetry for $0 < g < \infty$, so it can be used to construct the map from non-interacting states to hard-core repulsive states for multi-component fermions and bosons. More generally, this spectroscopic system holds promise for simplifying numerical methods for intermediate repulsive interaction strengths and attractive interactions, for perturbative approaches to systems with less symmetry, and for extensions to higher dimensions.

\section*{Acknowledgments}
Thanks to J.~DeMell, P.R.~Johnson, A.~Tsobanjan, and J.~Revels for numerous helpful discussions and especially to my students J. Verniero and B. Weinstein for their work on closely related projects. Also, the author appreciates the comments provided on an earlier draft by N.T.~Zinner, Z.~Zimboras, F.~Deuretzbacher and anonymous referees.

\section*{Appendix A: Reduction of $\mathrm{S}_N$ by subgroups}

To account for multicomponent bosons and fermions, irreps $\pp{p}$ of $\mathrm{S}_N$ can be reduced with respect to a subgroup and symmetrization rules applied at that level. For example, consider the case of $N$ two-component fermions where $N_1$ have one spin component (e.g.\ up)and $N_2$ have an orthogonal spin component (down). Then fermionic states will be in the one-dimensional irrep $\pp{1^{N_1}}\times \pp{1^{N_2}}$ of the subgroup $\mathrm{S}_{N_1}\times \mathrm{S}_{N_2}\subset \mathrm{S}_N$. Generally, each $\mathrm{S}_N$ irrep is a compound representation of a subgroup of $\mathrm{S}_N$. One can use the orthogonality relation between characters of a representation to find out whether $\pp{1^{N_1}}\times \pp{1^{N_2}}$ (or any other subgroup irrep) appears in the reduction of some particular $\mathrm{S}_N$ irrep. Tables \ref{tab:4red} summarizes these results for $N=4$, and Tables \ref{tab:3red} and \ref{tab:5red} show corresponding results for $N=3$ and $N=5$.

\begin{table}
\centering
\begin{tabular}{|c|c|ccc|}
\hline
number of & component & \multicolumn{3}{c|}{$\pp{p}$}\\
components &pattern &  \pp{3} & \pp{21} & \pp{1^3} \\
\hline
1 &$(3)_B$ &     1 & 0 & 0 \\
2 & $(21)_B$  &  1 & 1 & 0 \\ \hline
1 & $(3)_F$   &  0 & 0 & 1 \\
2 & $(21)_F$  &  0 & 1 & 1 \\ \hline
3 & $(111)$   & 1 & 2 & 1  \\
\hline
\end{tabular}
\caption{This table gives the reduction of $\mathrm{S}_3$ irreps by subgroups to account for multi-component bosons and fermions. The notation $(21)_B$ means two bosons in one component, and one bosons in another component. The case $(111)$ corresponds to distinguishable particles.}
\label{tab:3red}
\end{table}

\begin{table}
\centering
\begin{tabular}{|c|c|ccccccc|}
\hline
no.\ of & component & \multicolumn{7}{c|}{ $\pp{p}$}\\
comps.\ &pattern &  \pp{5} & \pp{41} & \pp{3^2} & \pp{31^2} & \pp{2^21} & \pp{21^3} & \pp{1^4} \\
\hline
1 & $(5)_B$ &    1 & 0 & 0 & 0 & 0 & 0 & 0 \\
2 & $(41)_B$  &  1 & 1 & 0 & 0 & 0 & 0 & 0\\
2 & $(32)_B$  &  1 & 1 & 1 & 0 & 0 & 0 & 0\\
2 & $(311)_B$  &  1 & 2 & 1 & 1 & 0 & 0 & 0\\
3 & $(221)_B$  &  1 & 2 & 2 & 1 & 1 & 0 & 0\\
4 & $(2111)_B$ &  1 & 3 & 3 & 3 & 2 & 1 & 0\\ \hline
1 & $(5)_F$ &     0 & 0 & 0 & 0 & 0 & 0 & 1 \\
2 & $(41)_F$  &  0 & 0 & 0 & 0 & 0 & 1 & 1\\
2 & $(32)_F$  &  0 & 0 & 0 & 0 & 1 & 1 & 1\\
2 & $(311)_F$  &  0 & 0 & 0 & 1 & 2 & 1 & 1\\
3 & $(221)_F$  &  0 & 0 & 1 & 1 & 2 & 2 & 1\\
4 & $(2111)_F$ &  0 & 1 & 2 & 3 & 3 & 3 & 1\\ \hline
5 & $(11111)$   & 1 & 4 & 5 & 6 & 5 & 4 & 1\\
\hline
\end{tabular}

\caption{This table gives the reduction of $\mathrm{S}_5$ irreps by subgroups to account for multi-component bosons and fermions. The notation $(221)_B$ means two bosons in one component, two bosons in another component and one in a third. The case $(11111)$ corresponds to distinguishable particles.}
\label{tab:5red}
\end{table}

\section*{Appendix B: Reduction of $\KHS^0_{\nu_R,\nu_\rho,\lambda}$}

The algorithm described in the main article is described in more detail here. To find the reduction of $\KHS^0_{\nu_R,\nu_\rho,\lambda}$, first reduce $\KHS^0_X$ into irreps of $\mathrm{S}_N$ using the symmetries of the partitions of $X$ into particle excitations. 
For example, when $N=4$ there are partitions with five types of symmetry:
\begin{itemize}
\item $(AAAA)$: subgroup $\mathrm{S}_4$ and multiplicity $1$,
\item $(AAAB)$: subgroup $\mathrm{S}_3$ and multiplicity $4$,
\item $(AABB)$: subgroup $\mathrm{S}_2\times\mathrm{S}_2$ and multiplicity $6$,
\item $(AABC)$: subgroup $\mathrm{S}_2$ and multiplicity $12$,
\item $(ABCD)$: subgroup $\mathrm{S}_1$ and multiplicity $24$.
\end{itemize}
Each $\KHS^0_X$ subspace can be decomposed into subspaces with basis vectors among those five types. As examples for $X<3$
\begin{itemize}
\item $X=0$: partition $(0,0,0,0)$
\item $X=1$: partitions $(1,0,0,0)$
\item $X=2$: partitions $(1,1,0,0)$, $(2,0,0,0)$
\item $X=3$: partitions $(1,1,1,0)$, $(2,1,0,0)$, $(3,0,0,0)$
\end{itemize}
Continuing with $X=3$ and $N=4$, there are four ways to distribute the energy partitions $(3,0,0,0)$ and $(1,1,1,0)$ over four particles and twelve ways to distribute $(2,1,0,0)$. Note that from Eq.~(\ref{dim}), $d^3_4 = 20 = 12 + 4+4$ as it should, i.e.\ the partitions $(3,0,0,0)$  and $(1,1,1,0)$ correspond to four-dimensional subspaces of $\KHS^0_3$ and $(2,1,0,0)$ to a 12-dimensional subspace of $\KHS^0_3$.

Each of these subspaces can be reduced into irreps of $\mathrm{S}_N$ by filling up standard Young tableaux with the available symbols. For example, the partition $(3,0,0,0)$ can be distributed in $\mathrm{S}_4$ standard Young tableaux as
\begin{equation}
\young(0003)\ \mbox{and}\ \young(000,3),
\end{equation}
the partition $(1,1,1,0)$ as
\begin{equation}
\young(0111)\ \mbox{and}\ \young(011,1),
\end{equation}
and the partition $(2,1,0,0)$ as
\begin{equation}
\young(0012)\ , \young(001,2),\ \young(002,1),\ \young(00,12)\ \mbox{and}\ \young(00,1,2).
\end{equation}
So this means for $N=4$, $\KHS^0_3$ can be reduced into three copies of $\pp{4}$, four copies of $\pp{31}$, one copy of $\pp{2^2}$, one copy of $\pp{21^2}$ and no copies of $\pp{1^4}$, or more briefly $\KHS^0_3 = \bigoplus [3,4,1,1,0]$. In a similar methods, one could find the following series for $N=4$: $\KHS^0_0 = \bigoplus [1,0,0,0,0]$, $\KHS^0_1 = \bigoplus [1,1,0,0,0]$ and $\KHS^0_2 = \bigoplus [2,2,1,0,0]$.

The subspace $\KHS^0_X$ can be also decomposed into the direct sum of subspaces $\KHS^0_{\nu_R,\nu_\rho,\lambda}$, each with dimensions $\epsilon_\lambda^N$. For example, $\KHS^0_1 = \KHS_{0,0,1}\oplus\KHS_{1,0,0}$, i.e.,\ one one-dimensional copy with $\lambda =0$ and one three-dimensional copy with $\lambda=1$. Combining these results with the above, then $\lambda=1$ subspaces correspond to the irrep pattern
\begin{eqnarray*}
\KHS^0_{0,0,1} &=& \KHS^0_1-\KHS^0_{1,0,0}\\
&=& \bigoplus[0,1,0,0,0],
\end{eqnarray*}
or one copy of $\pp{31}$. The $\KHS^0_2$ subspace has two copies of $\lambda=0$, one copy of $\lambda=1$ and one copy of $\lambda = 2$, so the five-dimensional $\lambda=2$ subspace can be reduced into
\begin{eqnarray*}
\KHS^0_{0,0,2} &=& \KHS^0_2 - \KHS^0_{2,0,0} - \KHS^0_{0,1,0} - \KHS^0_{1,0,1}\\
& =& \bigoplus[0,1,1,0,0],
\end{eqnarray*}
 i.e.\ $\pp{31}\oplus\pp{2^2}$. These results for $N=4$ are summarized in the main text, and the results for $N=3$ and $N=5$ are in Tables \ref{tab:gzero3} and \ref{tab:gzero5}.

\begin{table}
\centering
\begin{tabular}{|r|c|c|c|}
\hline
$\lambda$ & $\pp{3}$ & $\pp{21}$ & $\pp{1^3}$ \\
\hline
 0 & 1  & 0  & 0  \\
 1 & 0  & 1  & 0  \\
 2 & 0  & 1  & 0  \\
 3 & 1  & 0  & 1  \\
\hline
\end{tabular}
\caption{This table gives the degeneracy of $\mathrm{S}_3$ irreps in each $\KHS^0_{\nu_R,\nu_\rho,\lambda}$ subspace. For $\lambda>3$, the reductions follow the pattern $\KHS^0_{\nu_R,\nu_\rho,\lambda} = \KHS^0_{\nu_R,\nu_\rho,\lambda - 3}$}
\label{tab:gzero3}
\end{table}

\begin{table}
\centering
\begin{tabular}{|r|c|c|c|c|c|c|c|}
\hline
$\lambda$ & $\pp{5}$ & $\pp{41}$ & $\pp{32}$ & $\pp{31^2}$ & $\pp{2^21}$ & $\pp{21^3}$  & $\pp{1^5}$\\
\hline
 0 & 1  & 0  & 0  & 0  & 0  & 0  & 0\\
 1 & 0  & 1  & 0  & 0  & 0  & 0  & 0\\
 2 & 0  & 1  & 1  & 0  & 0  & 0  & 0\\
 3 & 1  & 1  & 1  & 1  & 0  & 0  & 0\\
 4 & 1  & 2  & 1  & 1  & 1  & 0  & 0\\
 5 & 1  & 2  & 2  & 2  & 1  & 0  & 0\\
 6 & 1  & 3  & 3  & 2  & 1  & 1  & 0\\
 7 & 1  & 4  & 3  & 3  & 2  & 1  & 0\\
 8 & 2  & 4  & 4  & 4  & 3  & 1  & 0\\
 9 & 2  & 5  & 5  & 5  & 3  & 2  & 0\\
10 & 2  & 6  & 6  & 6  & 4  & 2  & 1\\
11 & 2  & 7  & 7  & 7  & 5  & 3  & 0\\
12 & 3  & 8  & 8  & 8  & 6  & 4  & 0\\
13 & 3  & 9  & 9  & 10 & 7  & 4  & 1\\
\hline
\end{tabular}
\caption{This table gives the degeneracy of $\mathrm{S}_5$ irreps in each $\KHS^0_{\nu_R,\nu_\rho,\lambda}$ subspace.}
\label{tab:gzero5}
\end{table}

Combining these results with the calculations of Appendix A, degeneracy tables for $\KHS^0_{\nu_R,\nu_\rho,\lambda}$ and $\KHS^0_X$ are constructed for $N=3$, and $\KHS^0_{\nu_R,\nu_\rho,\lambda}$ for $N=5$ in Tables \ref{tab:3lam}-\ref{tab:5lam} below.

\begin{table}
\centering
\begin{tabular}{|c|ccccccc|}
\hline
component &  \multicolumn{7}{c|}{$\lambda$}\\
pattern &  0 & 1& 2& 3& 4 & 5 & 6  \\
\hline
$(3)_B$ &     1 & 0 & 0 & 1 & 0 & 0 & 1 \\
$(21)_B$ &    1 & 1 & 1 & 1 & 1 & 1 & 1 \\ \hline
$(3)_F$ &     0 & 0 & 0 & 1 & 0 & 0 & 1 \\
$(21)_F$ &    0 & 1 & 1 & 1 & 1 & 1 & 1 \\ \hline
$(111)$ &     1 & 2 & 2 & 2 & 2 & 2 & 2 \\
\hline
\end{tabular}
\caption{This table incorporates the results of Table \ref{tab:3red} to enumerate the degeneracy multi-component states obeying symmetrization rules in the space $\KHS^0_{\nu_R,\nu_\rho,\lambda}$ for $N=3$.}
\label{tab:3lam}
\end{table}

\begin{table}
\centering
\begin{tabular}{|c|ccccccc|}
\hline
component &  \multicolumn{7}{c|}{$X$}\\
pattern &  0 & 1& 2& 3& 4 & 5 & 6  \\
\hline
$(3)_B$ &     1 & 1 & 2 & 3 & 4 & 5 & 7 \\
$(21)_B$ &    1 & 2 & 4 & 6 & 9 & 12 & 16 \\ \hline
$(3)_F$ &     0 & 0 & 0 & 1 & 1 & 2 & 3 \\
$(21)_F$ &    0 & 1 & 2 & 4 & 6 & 9 & 12 \\ \hline
$(111)$ &     1 & 3 & 6 & 10 & 15 & 21 & 28 \\
\hline
\end{tabular}
\caption{This table incorporates the results of Table \ref{tab:3lam} to enumerate the degeneracy multi-component states obeying symmetrization rules in the space $\KHS^0_X$ for $N=3$.}
\label{tab:3X}
\end{table}

\begin{table}
\centering
\begin{tabular}{|c|ccccccccccc|}
\hline
component &  \multicolumn{11}{c|}{$\lambda$}\\
pattern &     0 & 1& 2& 3& 4 & 5 & 6 & 7 & 8 & 9 & 10 \\
\hline
$(5)_B$ &     1 & 0 & 0 & 1 & 1 & 1 & 1 & 1 & 2 & 2 & 2\\
$(41)_B$ &     1 & 1 & 1 & 2 & 3 & 3 & 4 & 5 & 6 & 7 & 8\\
$(32)_B$ &     1 & 1 & 2 & 3 & 4 & 5 & 7 & 8 & 10 & 12 & 14\\
$(311)_B$ &     1 & 2 & 3 & 5 & 7 & 9 & 12 & 15 & 18 & 22 & 26\\
$(221)_B$ &     1 & 2 & 4 & 6 & 9 & 12 & 16 & 20 & 25 & 30 & 36\\
$(2111)_B$ &     1 & 3 & 6 & 10 & 15 & 21 & 28 & 36 & 45 & 55 & 66\\ \hline
$(5)_F$ &     0 & 0 & 0 & 0 & 0 & 0 & 0 & 0 & 0 & 0 & 1\\
$(41)_F$ &     0 & 0 & 0 & 0 & 0 & 0 & 1 & 1 & 1 & 2 & 3\\
$(32)_F$ &     0 & 0 & 0 & 0 & 1 & 1 & 2 & 3 & 4 & 5 & 7\\
$(311)_F$ &     0 & 0 & 0 & 1 & 2 & 3 & 5 & 7 & 9 & 12 & 15\\
$(221)_F$ &     0 & 0 & 1 & 2 & 4 & 6 & 9 & 12 & 16 & 20 & 25\\
$(2111)_F$ &    0 & 1 & 3 & 6 & 10 & 15 & 21 & 28 & 36 & 45 & 55\\ \hline
$(11111)$ &     1 & 4 & 9 & 16 & 25 & 36 & 49 & 64 & 81 & 100 & 121\\
\hline
\end{tabular}
\caption{This table incorporates the results of Tables I and II in the main text to enumerate the degeneracy multi-component states obeying symmetrization rules in the space $\KHS^0_{\nu_R,\nu_\rho,\lambda}$ for $N=5$. Note that the degeneracy of $(11111)$ is $\epsilon^5_\lambda = (\lambda +1)^2$.}
\label{tab:5lam}
\end{table}

\section*{Appendix C: Reduction of $\KHS^\infty_{\nu_R,\nu_\rho,\lambda}$}

The character tables for $\mathrm{S_N}\times\mathrm{Z}_2$ can be derived from the properties of the transformations (3) and (4) in the main text. Note that all entries in the $N!\times N!$ matrices $\hat{U}(\gamma)$ are $1$'s, $-1$'s or $0$'s. The character of the identity is always the dimension of the representation $N!$. The character of all other pure  permutations is zero because all permutations map a function with support on a single sector to another sector so that have orthogonal domains. The character is also zero for the parity and for the parity times most permutations elements for the same reason. The only other non-zero characters are for elements $\pi$ and $c$ that reverse each other. For example, consider the element $c=(14)(23)$. It reverses the ordering  when  $\langle p \rangle = \langle 1234 \rangle$, $\langle 1324 \rangle$, $\langle 2143 \rangle$, $\langle 2413 \rangle$, $\langle 3142 \rangle$, $\langle 3412 \rangle$, $\langle 4231 \rangle$, or $\langle 4321 \rangle$. Parity inversion reverses this reversal and therefore, the character $\chi(\pi (14)(23))=8$ is, as is the character of the other two elements of that conjugacy class of $\mathrm{S}_N\otimes Z_2$. Character tables for irreps of $\mathrm{S_N}\times\mathrm{Z}_2$ and the $\hat{U}$ for $N=3$ and $N=4$ are in Tables \ref{tab:3char} and \ref{tab:4char}.

\begin{table}
\centering
\begin{tabular}{|c|cccccc|}
\hline
  & \multicolumn{6}{c|}{Conjugacy class}\\
Irrep & $\pp{1^3}$ & $\pp{21}$ & $\pp{3}$  & $i\pp{1^3}$ & $i\pp{21}$ & $i\pp{3}$\\
\hline
$\pp{3}^+$ &     1 & 1 & 1 & 1 & 1 & 1 \\
$\pp{21}^+$  &  2 & 0 & -1 & 2 & 0 & -1\\ 
$\pp{1^3}^+$   &  1 & -1 & 1 &  1 & -1 & 1 \\ 
$\pp{3}^-$ &     1 & 1 & 1 & -1 & -1 & -1 \\
$\pp{21}^-$  &  2 & 0 & -1 & -2 & 0 & 1\\ 
$\pp{1^3}^-$   &  1 & -1 & 1 &  -1 & 1 & -1\\ \hline
even $\lambda$  &  6 & 0 & 0 & 0 & -2 & 0\\
odd $\lambda$  &  6 & 0 & 0 & 0 & 2 & 0\\
\hline
\end{tabular}
\caption{This is the character table for $\mathrm{S}_3\times \mathrm{Z}_2$. The conjugacy classes are the normal conjugacy classes of $\mathrm{S}_3$ doubled by parity inversion $i$. The last two rows are the characters for the representation on the sector basis given by $\hat{U}(\gamma)$ described in the main text.}
\label{tab:3char}
\end{table}

\begin{table}
\centering
\begin{tabular}{|c|cccccccccc|}
\hline
  & \multicolumn{10}{c|}{Conjugacy class}\\
Irrep & $\pp{1^4}$ & \!\!\! $\pp{21^2}$ &  \!\!\!$\pp{2^2}$  & \!\!\! $\pp{31}$ &  \!\!\!$\pp{4}$  &  \!\!\!$i\pp{4}$ &  \!\!\!$i\pp{31}$ & \!\!\! $i\pp{2^2}$  &  \!\!\!$i\pp{21^2}$ & \!\!\! $i\pp{1^4}$\\
\hline
$\pp{4}^+$ &     1 & 1 & 1 & 1 & 1 & 1 & 1 & 1 & 1 & 1\\
$\pp{31}^+$   &  3 & 1 & -1 & 0 & -1 & 3 & 1 & -1 & 0 & -1\\
$\pp{2^2}^+$  &  2 & 0 & 2 & -1 & 0 & 2 & 0 & 2 & -1 & 0\\
$\pp{21^2}^+$   &  3 & -1 & -1 & 0 & 1 & 3 & -1 & -1 & 0 & 1\\
$\pp{1^4}^+$  & 1 & -1 & 1 & 1 & -1 & 1 & -1 & 1 & 1 & -1\\
$\pp{4}^-$ &    1 & 1 & 1 & 1 & 1 & 1 & 1 & 1 & 1 & 1 \\
$\pp{31}^-$   &  3 & 1 & -1 & 0 & -1 & -3 & -1 & 1 & 0 & 1\\
$\pp{2^2}^-$  & 2 & 0 & 2 & -1 & 0 & -2 & 0 & -2 & 1 & 0\\
$\pp{21^2}^-$   &  3 & -1 & -1 & 0 & 1 & -3 & 1 & 1 & 0 & -1\\
$\pp{1^4}^+$  & 1 & -1 & 1 & 1 & -1 & -1 & 1 & -1 & -1 & 1\\ \hline
even $\lambda$  &  24 & 0 & 0 & 0 & 0 & 0 & 0 & 8 & 0 & 0\\
odd $\lambda$  &  24 & 0 & 0 & 0 & 0 & 0 & 0 & -8 & 0 & 0\\
\hline
\end{tabular}
\caption{This is the character table for $\mathrm{S}_4\times \mathrm{Z}_2$. The conjugacy classes are the normal conjugacy classes of $\mathrm{S}_4$ doubled by parity inversion $i$. The last two rows are the characters for the representation on the sector basis given by $\hat{U}(\gamma)$ described in the main text.}
\label{tab:4char}
\end{table}

The reduction of $\KHS^\infty_{\nu_R,\nu_\rho,\lambda}$ is presented in the main text for $N=4$. Results for $N=3$ and $N=5$ are presented in Tables \ref{tab:ginfty3} and \ref{tab:ginfty5}.

\begin{table}
\centering
\begin{tabular}{|c|c|c|}
\hline
irrep    & even  &  odd \\
$\pp{p}^\pm$ &  $\lambda$ & $\lambda$\\
\hline
$\pp{3}^+$ & 0 & 1 \\
$\pp{21}^+$ & 1 & 1 \\
$\pp{1^3}^+$ & 1 & 0 \\
$\pp{3}^-$ & 1 & 0 \\
$\pp{21}^-$ & 1 & 1 \\
$\pp{1^3}^-$ & 0 & 1 \\
\hline
\end{tabular}
\caption{This table gives the reduction of the six-dimensional subspace $\KHS^\infty_{\nu_R,\nu_\rho,\lambda}$ into irreps of $\mathrm{S}_3\times \mathrm{Z}_2$ labeled as $\pp{p}^\pi$.}
\label{tab:ginfty3}
\end{table}

\begin{table}
\centering
\begin{tabular}{|c|c|c|}
\hline
irrep    & even  &  odd \\
$\pp{p}^\pm$ &  $\lambda$ & $\lambda$\\
\hline
$\pp{5}^+$ & 1 & 0 \\
$\pp{41}^+$ & 2 & 2 \\
$\pp{32}^+$ & 3 & 2 \\
$\pp{31^2}^+$ & 2 & 4 \\
$\pp{2^21}^+$ & 3 & 2 \\
$\pp{21^3}^+$ & 2 & 2 \\
$\pp{1^5}^+$ & 1 & 0 \\
$\pp{5}^-$ & 0 & 1 \\
$\pp{41}^-$ & 2 & 2 \\
$\pp{32}^-$ & 2 & 3 \\
$\pp{31^2}^-$ & 4 & 2 \\
$\pp{2^21}^-$ & 2 & 3 \\
$\pp{21^3}^-$ & 2 & 2 \\
$\pp{1^5}^-$ & 0 & 1 \\\hline
\end{tabular}
\caption{This table gives the reduction of the $120$-dimensional subspace $\KHS^\infty_{\nu_R,\nu_\rho,\lambda}$ into irreps of $\mathrm{S}_5\times \mathrm{Z}_2$ labeled as $\pp{p}^\pi$.}
\label{tab:ginfty5}
\end{table}

\end{document}